**Measurement Anomaly of Step Width in Calibration Grating using Atomic Force Microscopy**

Gun Ahn,[1,2] Yoon-Young Choi,[1] Dean J. Miller,[3] Hanwook Song,[4] Kwangsoo No,[2,*] and Seungbum Hong[1,2,*]

[1]*Materials Science Division, Argonne National Laboratory, Lemont, IL*

[2]*Department of Materials Science and Engineering, KAIST, 291 Daehak-ro, Yuseong-gu, Daejeon, 305-701, South Korea*

[3]*Nanoscience and Technology Division, Argonne National Laboratory, Lemont, IL*

[4]*Center for Mass and Related Quantities, Division of Physical Metrology, Korea Research Institute of Standards and Science, 267 Gajeong-ro, Yuseong-gu, Daejeon 305-340, South Korea*

[*]*Corresponding authors*

*Seungbum Hong*

Phone number: +82-42-350-3324

Fax number: +82-42-350-3310

E-mail: seungbum@kaist.ac.kr

*Kwangsoo No*

Phone number: +82-42-350-3370

Fax number: +82-42-350-3310

E-mail: ksno@kaist.ac.kr

**Abstract**

We imaged the topography of a silicon grating with atomic force microscopy (AFM) using different scan parameters to probe the effect of pixel pitch on resolution. We found variations in the measured step height and profile of the grating depending on scan parameters, with measured step width decreasing from 1300 to 108 nm and step height increasing from 172 to 184 nm when a pixel pitch in the scan axis decreased from 625 nm to 3.91 nm. In order to resolve the measurement anomaly of step width, we compared these values with step width and height of the same grating measured using scanning electron microscopy (SEM). The values obtained from SEM imaging were 187.3 nm ± 6.2 nm and 116 nm ± 10.4 nm, which were in good agreement with AFM data using a 3.91 nm of pixel pitch. We think that we need at least four pixels over the step width to avoid the measurement anomaly induced by the stick-slip or dragging of the tip. Our findings that RMS roughness varied less than 1 nm and converged at the value of 77.6 nm for any pixel pitch suggest that the RMS roughness is relatively insensitive to the pixel pitch.

**Introduction**

In the semiconductor industry, recessed structures are often adopted to increase the surface area of electrodes for capacitors [Park et al., 2014]. As such, knowing the exact dimension of a sub-micron scale trench is of great importance. For capacitors, the interface roughness determines the reliability due to the lightning rod effect of nanoscale protruding region [Choi et al., 2005].

Up to now, both top view and cross-section scanning electron microscopy (SEM) have been successfully utilized for this purpose in the semiconductor industry [Kato et al., 2014]. However, as the critical dimension shrinks to tens of nanometers, the measurement becomes more challenging [Grabar et al., 1997]. Moreover, quantitative measurement of surface roughness is very difficult using SEM [Smiley et al., 1993].

Unlike SEM, which provides a two-dimensional projection of a three-dimensional surface of a sample, atomic force microscopy (AFM) provides a three-dimensional surface profile. Additionally, samples viewed by AFM do not require any special sample preparations [Binnig et al., 1986]. Thus, AFM is viewed as an attractive approach to evaluate the surface topography and dimensions of sub-micron features in semiconductor devices.

However, it has often been questioned why the surface morphology observed by AFM sometimes differ from that obtained by SEM and whether AFM scan parameters may influence the results. These questions motivated us to conduct the work presented here. In order to simplify the experiment but still address the issue of discrepancy between measurement approaches, we used a simple calibration grating as our model system and measured the horizontal and vertical dimensions of the grating trenches as the basic component of morphology. We varied the number of total pixels, scan size and scan rate to study the effect of those parameters in the morphology obtained by AFM and identified the optimum parameters to obtain the same morphology as obtained by SEM.

**Methods**

We used a commercial AFM (MFP-3D, Asylum research) and Pt coated silicon cantilevers (EFM tip, Nanoworld) in our experiments. We used a silicon-grating sample (CalibratAR 3D Calibration Reference for X and Y are 10 μm +/- 0.04 μm, Z depth is 200nm +/- 4nm calibration of the scanning mechanism, Asylum Research) as our model sample.

The sample was scanned ten times at different pixel pitch of 3.91, 19.53, 23.44, 31.25, 39.06, 78.13, 156.25, 312.5 and 625 nm. We achieved those pitches by varying either the scan size from 1 μm × 1 μm to 40 μm × 40 μm or the number of pixels from 64 × 64 to 1024 × 1024. We fixed the scan rate to be 1 Hz and used scan angles of 0° and 45°. The AFM tip was grounded to prevent static electricity while scanning the sample.

We chose ten arbitrary lines perpendicular to the boundary between the surface and the trench of the silicon grating in each image, and measured the average step width and height using the line profiles. We also measured the root-mean-square (RMS) of the sample with different pixel pitch over 40 μm × 40 μm for 10 times.

Finally, we measured the step height and width of the silicon grating for 10 grating steps using scanning electron microscopy (SEM) after etching part of the trench by focused ion beam (FIB) milling.

**Results**

Figure 1 shows the topography images and line profiles along an arbitrarily chosen cross - section of the silicon grating collected using different pixel pitch defined by the scan size divided by the total number of the pixels in each image. The boundary edges between the square holes and the flat region become straighter and sharper as the pixel pitch decreases as expected. Small structural details become visible as the pixel pitch decreases. If we compare the topography images or the line profile, by eye, one cannot see much difference in the resolution between 256 × 256 pixels and 1,024 × 1,024 pixels. For more detailed analysis, we measured the step width and height of the Si grating.

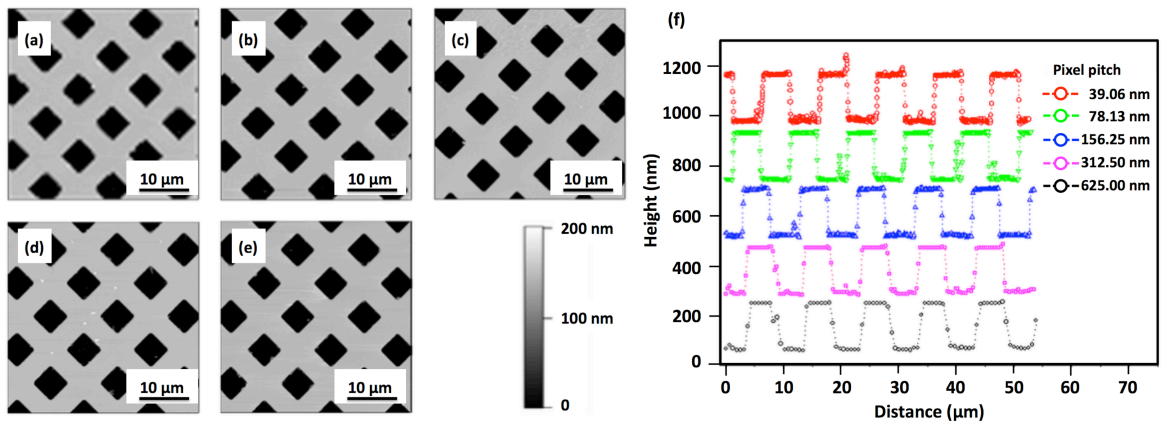

**Figure 1**. The topography images of the silicon grating with different pixel pitch: (a) 625 nm, (b) 312.5 nm, (c) 156.3 nm, (d) 78.1 nm and (e) 39.1 nm. The scan size is 40 × 40 μm and total numbers of pixels are (a) 64 × 64, (b) 128 × 128, (c) 256 × 256, (d) 512 × 512, and (e) 1,024 × 1,024, respectively. (f) The line profile along arbitrarily chosen cross-section of the silicon grating with different pixel pitch.

Figure 2(a) shows the step width and height as a function of pixel pitch of the grating sample. The measured step width decreased from 1300 nm to 310 nm and the measured step height increased from 171.8 nm to 183.8 nm until the pixel pitch reached 78.13 nm. We found that the step height measured at the pixel pitch of 78.13 nm was the same within the confidence level of 5 % as those measured at the pixel pitch of 39.06 nm based on our analysis of variance (ANOVA). However, the step width and height measured by the SEM image (Figure 2(b)) were 115.9 ± 10.4 nm and 187.3 ± 6.2 nm, respectively.

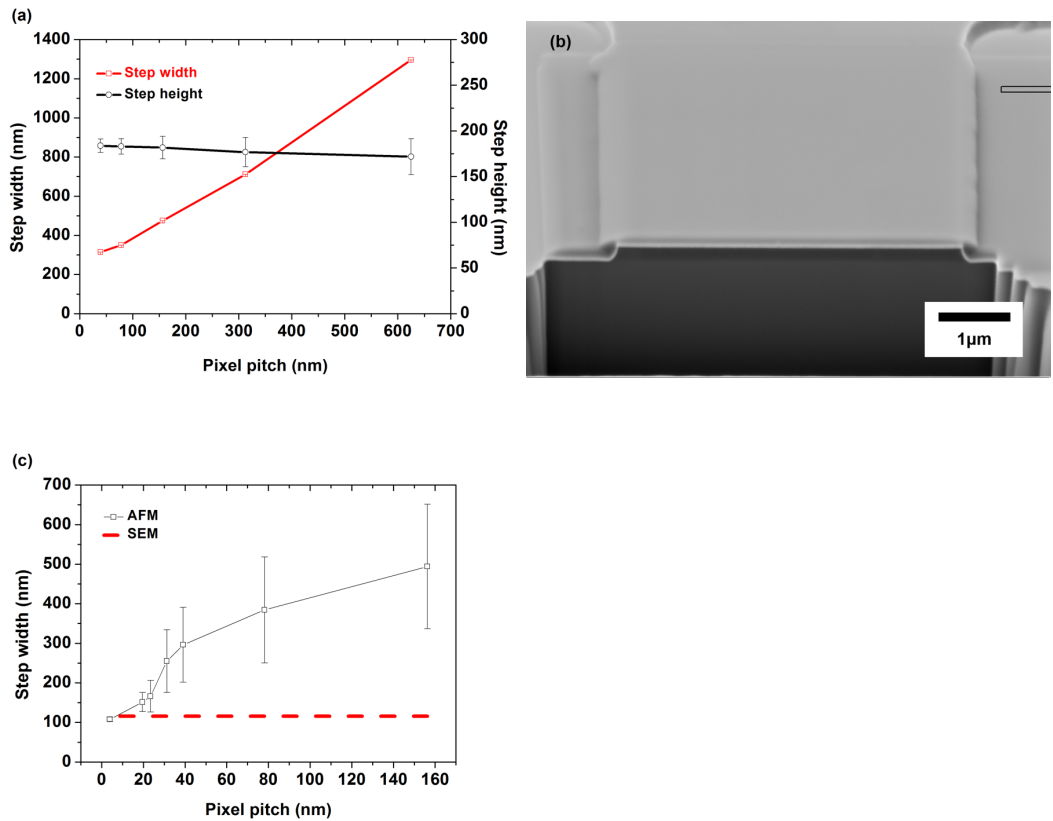

**Figure 2**. (a) The step width and height as a function of pixel pitch of the grating sample. (b) Scanning Electron Microscopy (SEM) image of silicon grating etched by Focused Ion Beam (FIB) milling. (c) The step width as a function of pixel pitch of the grating sample. Pixel pitch is smaller than that in Figure 2 (a).

There are another important parameters to analyze when we image the surface of material. In two dimensional surfaces, the flatness of surface is an important issue. To measure the flatness and its accuracy, we measured root mean square (RMS) roughness of the silicon grating.

Figure 3 shows the root mean square (RMS) roughness as a function of pixel pitch. It is found that RMS roughness converged to the value of 77.6 nm as a function of pixel pitch of number of pixels.

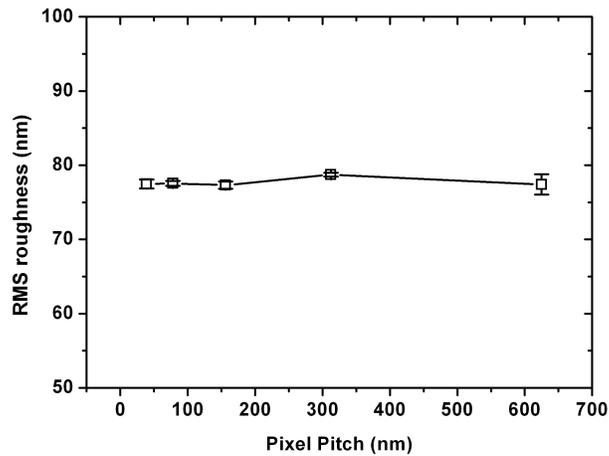

**Figure 3**. The plot of root mean square (RMS) roughness as a function of pixel pitch.

**Discussion**

Based on our measurement with AFM, the step height data was reliable but the step width was not. There are several reasons behind artifacts responsible for the discrepancy of step width measured by AFM when compared with the one measured by SEM. The AFM tip radius, tilt correction, and scanner nonlinearity can play significant role in the artifacts [Gavrilenko et al., 2009]. The radius and height of tip were 30 nm and 12.5 μm, respectively, which may explain a discrepancy up to 30 nm. However, the discrepancy in step width was 71.4 nm in our experiments.

Therefore, we investigated the effect of pixel pitch and used smaller scan size to achieve both smaller pixel pitch and less scanner nonlinearity. We found that the measured step width converged to the value obtained by SEM when the pixel pitch was 3.91 nm as shown in Figure 2(c). We are tempted to attribute this finding to non-linearity of our scanner. However, this alone cannot account for the fact that we need at least 30 points for 1 μm scan to measure the step width of about 116 nm. One more potential contributor to this phenomenon might be the stick-slip or dragging of the tip, which is in contact with the surface [Choi et al., 2011]. It should be noted that the stick-slip or dragging of the tip depends on the sample's surface condition such as roughness, hydrophilicity/hydrophobicity or capillary force induced by the water meniscus between the AFM tip and the surface. Error in acquisition by one or two or three pixels will result in the discrepancy proportional to the pixel pitch. This is indeed the case in Figure 2(c).

Based on our analysis of variance (ANOVA), we found that the RMS values did not change as a function of pixel pitch within a 5% confidence level. It is somewhat unexpected to observe no significant change in RMS roughness as a function of pixel pitch as we found strong dependence of step width on the pixel pitch. This can be attributed to the fact that change in one or two pixels does not contribute to the overall roughness. There is an experimental advantage when we use larger pixel pitch for obtaining the roughness as it takes less time to acquire the full image.

In conclusion, we identified optimum resolution for fast and reliable topography acquisition using atomic force microscopy (AFM). We imaged the topography of a silicon grating using different pixel pitches and found that the measured step width decreased from 1300 to 108 nm and the step height increased from 172 to 184 nm when the pixel pitch in the fast scan axis decreased from 625 nm to 3.9 nm. Using scanning electron microscopy (SEM), we also measured the step width and height of the silicon grating in cross-sections revealed by focused ion beam (FIB) milling, and compared the measured step height and width to the AFM data. The values obtained from SEM images were 187.3 nm ± 6.2 nm and 116 nm ± 10.4 nm, which are in agreement with AFM data collected with less than 39 nm and 23 nm of pixel pitch, respectively. Lastly, our findings that RMS roughness varied less than 1 nm and converged at the value of 77.6 nm with any pixel pitch suggest that we can reliably use the RMS roughness obtained over a wide range of scan resolutions. Based on the assumption that our results scale linearly with the size of the features of interest, we believe that one can use the optimal pixel pitch for fast and reliable topography acquisition and surface roughness analysis.

**Acknowledgement**

This work was supported by the U.S. Department of Energy, Office of Science, Materials Sciences and Engineering Division. The FIB and SEM was carried out in the Electron Microscopy Center, a U.S. Department of Energy, Office of Science, Office of Basic Energy Sciences User Facilities, under Contract DE-AC02-06CH11357. K. N. acknowledges financial support from Mid-career researcher program (No. 2010-0015063) and conversion research center program (No.2011K000674) through the National Research Foundation (NRF) funded by the Ministry of Education, Korea.